\documentclass[11pt,aps,nofootinbib,prd,aps,epsf,floats,axodraw,amsmath,amssymb,amsfonts]{revtex4}
\usepackage{amsmath, amssymb}
\newcommand{\mathsym}[1]{{}}

\usepackage{graphicx}
\usepackage{amsmath}
\usepackage{amssymb}
\usepackage{bm}
\newcommand{\be}{\begin{equation}}
\newcommand{\ee}{\end{equation}}
\newcommand{\bea}{\begin{eqnarray}}
\newcommand{\eea}{\end{eqnarray}}

\newcommand{\rem}[1]{}
\newsavebox{\PSLASH}
 \sbox{\PSLASH}{$p$\hspace{-1.8mm}/}
 
\renewcommand{\theequation}{\thesection.\arabic{equation}}
\newcounter{saveeqn}
\newcommand{\add}{\addtocounter{equation}{1}}
\newcommand{\alpheqn}{\setcounter{saveeqn}{\value{equation}}%
\setcounter{equation}{0}%
\renewcommand{\theequation}{\mbox{\thesection.\arabic{saveeqn}{\alph{equation}}}}}
\newcommand{\reseteqn}{\setcounter{equation}{\value{saveeqn}}%
\renewcommand{\theequation}{\thesection.\arabic{equation}}}

 \newsavebox{\notrightarrow}
 \sbox{\notrightarrow}{$\to$\hspace{-4mm}/}
 
 \newsavebox{\PARTIALSLASH}
 \sbox{\PARTIALSLASH}{$\partial$\hspace{-1.6mm}/}
 
 \newsavebox{\ASLASH}
 \sbox{\ASLASH}{$A$\hspace{-2.1mm}/}
 
 \newsavebox{\KSLASH}
 \sbox{\KSLASH}{$k$\hspace{-1.8mm}/}
 
 \newsavebox{\LSLASH}
 \sbox{\LSLASH}{$\ell$\hspace{-1.8mm}/}
 
 \newsavebox{\QSLASH}
 \sbox{\QSLASH}{$q$\hspace{-1.8mm}/}
 
 \newsavebox{\DSLASH}
 \sbox{\DSLASH}{$D$\hspace{-2.2mm}/}
 
 \newsavebox{\DbfSLASH}
 \sbox{\DbfSLASH}{${\mathbf D}$\hspace{-2.8mm}/}
 
 \newsavebox{\DELVECRIGHT}
 \sbox{\DELVECRIGHT}{$\stackrel{\rightarrow}{\partial}$}
 
 \newcommand{\blue}{\IfColor{\textCadetBlue}{}}
\newcommand{\black}{\IfColor{\textBlack}{}}
\newcommand{\red}{\IfColor{\textRed}{}}
\newcommand{\green}{\IfColor{\textOliveGreen}{}}
\newcommand{\lila}{\IfColor{\textRedViolet}{}}

\begin{document}
\begin{flushright}
 [hep-th/math-ph/quant-ph]
\end{flushright}
\title{Brownian Motion of the Quantum States on a String and the Polyakov Action of String Theory:\\ Is String Theory a Quantum Mechanical Model of the Brain?}

\author{Amir Abbass Varshovi}


\begin{abstract}
   \textbf{Abstract\textbf{:}} The Brownian motion of a number of quantum states in a compact one-dimensional space is studied via the Wiener fractal measure, and it is shown that the derived path-integral measure coincides precisely with the Polyakov path-integral formula for bosonic string theory. Thus, it is concluded that the Polyakov action of bosonic string theory does not have a unique, distinguishable foundation specifically dedicated to describing the fundamental forces of nature, but rather, it is merely a standard formulation of the Wiener stochastic process for Brownian motion of the quantum states on one-dimensional objects. It is also demonstrated that the time-dimension field $X^0$ is, in practice, the localization of the non-local effects of the coordinate fields $X^i$s, $1\leq i \leq D-1$. This indicates that the interpretation intended for spacetime fields in the formulation of string theory allegedly faces fundamental flaws in its underlying theoretical aspects. In this regard, we will defend string theory against its experimental flaws in particle physics due to unreliable interpretation of the theory and relate its elaborated mathematical framework to another significant topic: Quantum Brain. It is then argued that the formulation of bosonic string theory can be applied to the quantum behavior of any microscopic filament, such as those found in cellular substructures and neuronal subunits. We conclude that if a neuronal subunit can be found encompassing $D-1=25$ real quantum states, it can be considered a sustainable quantum system with a consistent bosonic string theory in $D=26$ dimensions. Fortunately, such a neuronal subunit had already been recognized in distinct studies by Penrose and Hameroff as the origin of brain quantum activities: Microtubule. Microtubules are composed of 13 protofilaments, collectively generating 25 independent real quantum states, producing a well-defined string-theoretic quantum system in neurons. The results of our study are compared with the achievements of the Orch-OR theory, and remarkable consistencies between the two approaches are presented. Finally, the non-locality field $X^0$ is recognized for both the time dimension in the background mathematics of string theory on the one hand and the source of consciousness in quantum neurobiology on the other hand. This conformity is in complete agreement with Hameroff's recent hypothesis on time-consciousness identification. \\
       
\noindent \textbf{Keywords\textbf{:}} Weierstrass-like Functions, Fractal Geometry, Fractal Norm, Wiener Stochastic Process, Brownian Motion, Wiener Fractal Measure, String Theory, Bosonic Strings, Non-Locality Field, Neurons, Quantum Neurobiology, Quantum Mind, Quantum Model of Brain, Microtubule, Protofilament, Orch-OR Theory, Consciousness Field, Flow of Time.
\end{abstract}

 \maketitle

\newpage
\tableofcontents 

~\\\\

\section{Introduction}
\label{introduction}

\par The exploration of the brain's functioning through the lens of quantum mechanics has given rise to intriguing interdisciplinary fields of study such as quantum neurobiology and the quantum mind hypothesis.\footnote{See \cite{Penrose, Hameroff2020, Hameroff2022, Hameroff2024, Ham-Pen1996a, Penrose1995, Hameroff2014, Hameroff2014a, Hameroff2003, Hameroff2012, Penrose2014, Jedlicka, Roth, Dey, Kher, Donald, Geor, Nan, Koch, Schwartz, Swan} for more discussions.} These fields investigate how quantum phenomena might underlie cognitive processes and the fundamental features of consciousness. Motivated by these ideas, this paper first examines the Brownian motion of quantum states on strings within the corresponding infinite-dimensional Hilbert space. This provides us with a comprehensive understanding of the quantum mechanical dynamics within neurons or their subunits having string-like molecular structures. Subsequently, based on the results of the aforementioned perusal we explore the correlation between string theory and the quantum mechanisms governing brain function.

\par Recent advances in quantum mechanics and neurobiology suggest that the brain operates in ways that might be deeply rooted in quantum processes, such as superposition and entanglement.\footnote{See \cite{Jedlicka, Roth, Dey, Kher, Donald, Geor, Nan, Koch, Schwartz, Swan, Ham-Pen1996a, Penrose1995} and the references therein for more information.} String theory, with its profound implications for understanding the fundamental structure of reality, offers a novel perspective that could illuminate the complexities of brain function at the quantum level. This paper will examine the intersection of these domains, discussing how principles from string theory might inform our understanding of quantum neurobiology and the quantum mind, and potentially unravel new dimensions of human cognition and consciousness.

\par In our recent work \cite{Varshovi1},\footnote{Accompanied with the subsequent preprint work \cite{Varshovi2}.} we employed the Wiener stochastic process of Brownian motions via the corresponding path-integral formulation on the one hand, and utilized the Wilsonian renormalization approach based on the fractal infrastructures of quantum states at high energy levels on the other hand to study an essentially basic stochastic problem in quantum physics: The Brownian motion of quantum states on a closed Riemannian manifold.\footnote{In \cite{Varshovi2} we have proven that the conclusion is valid in the presence of non-trivial Riemannian metrics and background gravitational effects.} We observed there that the Wiener path-integral measure for the mentioned theoretical ingredients and structures, the so-called \emph{Wiener fractal measure}, astonishingly coincides with Feynman's path-integral measure for scalar quantum fields.\footnote{Furthermore, in our subsequent study \cite{Varshovi3}, we have shown that this procedure can be extended to gauge and spinor fields and have again observed that the Wiener fractal measure is practically equal to Feynman's path integral measure for Yang-Mills fields even in the presence of gravity.} As such, this thought-provoking question arises: If Brownian motion of quantum states can produce the foundational formulations of quantum field theory for elementary particles, are they capable of also creating the mathematical framework of string theory?

\par There is no need for explanation that string theory is undoubtedly the most complete, coherent, and yet most beautiful mathematical formulation of our physical assumptions. However, despite this, the theory suffers significantly from major deficiencies in the realm of experimental studies, since, in fact, it has practically offered no definite and clear description of the interaction of elementary particles.\footnote{See \cite{Smolin, Woit, Dawid, Ritson, Rovelli, Rovelli1998, Rovelli2003, Rovelli2016, Conlon} for more discussions.} This problem draws attention to the question of whether string theory is fundamentally talking about distinct objects and subjects rather than fundamental ingredients of the quantum infrastructures of spacetime and the foundational elements of natural forces. 

\par Responding to the question of what string theory is essentially talking about is a highly complex task unless we can demonstrate that some quantum systems, which inherently cannot be identified with the basic systems of fundamental forces, such as some microscopic molecular stringy objects with quantum mechanical effects, also adhere to the equations of string theory. In that case, not only would we have concrete examples for practical study and experimentation of string theory, but we would also be able to achieve more precise interpretations of the fundamental elements of string theory, such as spacetime fields $X^\mu$s. In this regard, without any prejudice, we should anticipate unexpected interpretations of the contents and ingredients of string theory and its solutions.

\par In this paper, by employing the Wiener fractal measure for Brownian motion of quantum states we will show that any microscopic string-like object that can contain 25 real quantum states (or 13 complex Schrodinger's wavefunctions with an analytic or topological constraint) can be considered as a sustainable quantum system whose quantum dynamics and interactions can be accurately described by the formulations of the well-defined bosonic string theory in $D=26$ dimensions. Obviously, finding such attractive systems in nature is very difficult; however, fabricating them in quantum physics laboratories appears possible. One such unique object in nature is the extremely mysterious subunit of cells: the microtubule. 

\par Microtubules are cellular components crucial for cell division, characterized by some periodic lattice structures.\footnote{See \cite{Cooper} for more information about microtubules.} They consist of 13 protofilaments, each containing two real quantum states (real and imaginary parts of the Schrodinger wavefunction), resulting in 25 independent real quantum states due to the underlying topological structures. Consequently, microtubules will consistently exhibit sustainable string-theoretic effects and activities in $D=26$ bosonic dimensions. With this discovery, our research will focus on understanding the performance of this string-theoretic quantum machine in brain neurons. 

\par Penrose and Hameroff have speculated that microtubules are responsible for the brain's quantum mechanical effects, particularly the non-local aspects of consciousness.\footnote{See \cite{Penrose, Hameroff2020, Hameroff2022, Hameroff2024, Ham-Pen1996a, Penrose1995, Hameroff2014, Hameroff2014a, Hameroff2003, Hameroff2012, Penrose2014} for more details.} The surprising point in our calculations is that we prove that all quantum non-local processes are included in the string theory formulation within a local mathematical framework. Thus, the quantum performance of microtubules based on the Polyakov path-integral includes quantum non-local interactions. This fact supports Penrose and Hameroff's theory about the correlation of consciousness and non-local interactions in the quantum mechanical activities of the brain.\footnote{See \cite{Penrose, Hameroff2020, Hameroff2022, Hameroff2012, Hameroff2014a, Hameroff2024, Herce} for more details.}

\par In the first part of this article (Section II), we will provide a brief review of the extraction process for the Wiener fractal measure for bosonic states. Following this, we will work out the path-integral measure for the Brownian motion of $D-1$ independent real (bosonic) quantum states on a string. We will see that this process, after discarding the whole non-local effects, will produce the Polyakov action in $D-1$-dimensional Euclidean spacetime. In the subsequent section (Section III), we endeavor to reintroduce the non-local effects into the formulation by incorporating an augmented local field $X^0$, the so-called \emph{non-locality field}. Upon doing so, we will observe that upon reactivating the non-local effects (with the utmost precision), the Wiener fractal measure aligns with the Polyakov path-integral measure for bosonic string theory in $D$-dimensional Lorentzian spacetime.

\par In light of our precise extraction of the Polyakov path-integral formulation of bosonic strings and our distinct interpretation of the $X^0$ field as representing non-localities, skepticism arises concerning the validity of the prevailing interpretations regarding the theoretical foundations and achievements of string theory. Thus, in the next section (Section IV), we try to provide a list of criticisms of string theory regarding its lack of refutability and experimental verifiability, and then open new doors to the problem in order to defend string theory and convey the mentioned flaws to the misunderstanding of the theory and its underlying fundamental concepts, elements, and ingredients.

\par In the concluding section (Section V), we contend that irrespective of the validity of string theory's applicability in exploring fundamental forces of nature, it holds relevance in understanding quantum neurobiological aspects and the concept of the quantum brain. We then introduce microtubules and explain why they can be a perfect and sustainable example for generating a consistent string theory. This is the moment when we surprisingly encounter Penrose and Hameroff's Orch-OR theory. Subsequently, we will discern multiple analogous findings associated with this theory, albeit derived from distinct and diverse methodologies.

\par In summary, the article endeavors to address two pivotal inquiries:\\

\par \textbf{Question 1)} \emph{What path-integral measure formulation describes the Brownian motion of quantum states of strings using the Wiener fractal process?}\\

\par \textbf{Question 2)} \emph{What quantum theory addresses the quantum activities of the brain and can explain its quantum mechanical phenomena, such as consciousness?} \\

\par This paper asserts that both questions share a common answer: the \emph{Bosonic String Theory.}


\par
\section{Review: Formulating the Brownian Motion of Quantum States}
\setcounter{equation}{0}

\par Brownian motion, in its inherent mathematical essence, describes the random movement of general objects in a medium space. The Wiener process formulates Brownian motion in continuous-time stochastic processes, providing the basis for stochastic calculus and serving as a fundamental tool in studying processes with inherent uncertainty.\footnote{See \cite{Folland, Durret}} Its importance lies in its rigorous measure-theoretic (probabilistic) framework to analyze and understand randomness and uncertainty,\footnote{See \cite{Wiener1, Wiener2}.} enabling the stochastic formulation of quantum mechanical behaviors of quantum objects including elementary particles like the Higgs boson \cite{Varshovi1}.

\par Let us study the Brownian motion of the quantum states on a string with length $\ell$. Indeed, in this case, any quantum state is generally a continuous function $f:I \to \mathbb C$ satisfying the property
\begin{equation} \label {normal}
\int_I |f(\sigma)|^2 d\sigma = 1,
\end{equation}

\noindent wherein $I=[0,\ell]$ and $\sigma$ is the length variable along the string. According to our main purpose in this study, without loss of generality, we can assume that $f$ is a real function at the cost of relaxing the relation (\ref{normal}).\footnote{See \cite{Varshovi1} for similar assumptions and the mathematical mechanism due. One should note that the ultimate formulation of our perusal will include complex quantum states as a theory for two real quantum states. However, the Brownian motion of real quantum states leads to the scalar, Yang-Mills, and spinor quantum field theories together with the gravitational effects \cite{Varshovi1, Varshovi2, Varshovi3} which include most structures of the Standard Model and beyond. Moreover, here we aim to study the fundamental machinery of bosonic string theory which is, in fact, a theoretical model for real quantum states on strings.} Substantially, such an assumption lets us assume each Schrodinger wavefunction as two real quantum states, while in practice the imposed (topological or geometric) constraints can affect the number of independent quantum mechanical ingredients. 

\par Therefore, our study concerns with the Brownian motion of continuous real functions on the string $f:I \to \mathbb R$ along the flow of time. Regardless of the boundary conditions, each such function, if differentiable, has a specific Fourier expansion based on the Laplacian eigenfunctions $\cos(2n\pi\sigma / \ell)$ and $\sin(2n\pi\sigma / \ell)$ for $n\geq 0$. In the following, we show the Fourier coefficients of $f$ with $f_n$, where $n$ is any number of $\mathbb Z$ so that the cosine functions $\cos(2n\pi\sigma/ \ell)$ correspond to non-negative integers and the sine functions $\sin(2n\pi\sigma/ \ell)$ are attached to negative Fourier modes. Thus, we have:
\begin{equation}
f_n=\left\{ {\begin{array}{*{20}{c}}
   {\frac{2}{\ell}\int_I f(\sigma)\cos(2n\pi\sigma / \ell) d\sigma,~~~~~~~~~~~~n>0}  \\
   {\frac{1}{\ell}\int_I f(\sigma) d\sigma},~~~~~~~~~~~~~~~~~~~~~~~~~~~~n=0\\
   {\frac{2}{\ell}\int_I f(\sigma)\sin(2|n|\pi\sigma / \ell) d\sigma},~~~~~~~~~~n<0  \\
\end{array}} \right.  ,
\end{equation}

\noindent with the simple property of
\begin{equation} \label{Derivative}
f'_n = \frac{2\pi}{\ell} n f_{-n},
\end{equation}

\noindent for $f'(\sigma ) = d f(\sigma )/d\sigma$.

\par However, there are continuous functions $f:I \to \mathbb R$ which admit Fourier expansion but are not differentiable on a dense subset of $I$. Such functions were first discovered by Karl Theodor Wilhelm Weierstrass in his seminal research in analysis \cite{Weierstrass} which introduced the celebrated Weierstrass function and hence, we mostly refer to them as \emph{Weierstrass-like} or \emph{fractal functions}. Fractal functions have been studied thoroughly in our previous work \cite{Varshovi1} and we proved that their Fourier coefficients $f_n$ obey a specific asymptotic property. More precisely, based on the achievements of Weierstrass \cite{Weierstrass}, and Hardy \cite{Hardy}, we have established that there is an asymptotic rule for $f_n$s which can single out the fractal functions. The rule has been stated in the following theorem:\\

\textbf{Theorem 1 \cite{Varshovi1};} \emph{Assume that $I \subset \mathbb{R}$ is a compact interval and $f(x):I \to \mathbb{R}$ is a continuous function with Fourier coefficients $f_n$. Then, $f(x)$ is a Weierstrass-like function on the whole domain $I$ if and only if for any $\kappa>0$ and $l_0>0$, there exists a Fourier mode $n$ so that the inequality}
\begin{equation} \label {Th1}
\int_0^{|f_n|}~e^{\kappa n^2 x^2/2}~dx > l_0
\end{equation}

\noindent \emph{holds for the Fourier coefficient $f_n$.}\\

\par The above theorem plays a pivotal role in studying the Brownian motion of quantum states in the corresponding infinite-dimensional Hilbert space. It is well-known that the Brownian motion of a particle in $\mathcal D$-dimensional medium space is given through the celebrated Wiener probability measure for stochastic processes. Strictly speaking, if the particle is located at $x_0 \in \mathbb R^{\mathcal D}$ at $t_0$, then the probability of finding the particle in $U_i \subset \mathbb R^{\mathcal D}$ at $t_i$, $1\leq i \leq N$, with the ordering of $t_0<t_1<\cdots <t_N$ is given via the following formula \cite{Wiener1, Wiener2};\footnote{See also \cite{Folland, Durret} as two beautiful and comprehensive presentations of the issue.}
\begin{equation} \label {PU}
\begin{gathered}
\mathcal P(U_1,\cdots, U_N;t_1,\cdots,t_N)
=\left( \frac{1}{2\pi \Delta t_N}\right)^{\mathcal D/2} \cdots \left( \frac{1}{2\pi \Delta t_1} \right)^{\mathcal D/2} \\ \times \int_{U_N}\cdots \int_{U_1} e^{-|x_N-x_{N-1}|^2/2\Delta t_N}\cdots e^{-|x_2-x_1|^2/2\Delta t_2} e^{-|x_1-x_0|^2/2\Delta t_1}~d^{\mathcal D}x_N \cdots d^{\mathcal D} x_1,
\end{gathered}
\end{equation}

\noindent wherein $\Delta t_i=t_i-t_{i-1}$, $1\leq i\leq N$. Also, we used $x_i=(x_i^1,\cdots,x_i^{\mathcal D})$ and $d^{\mathcal D}x_i$ for the coordinate system and the Lebesgue measure of $\mathbb R^{\mathcal D}$ at $t_i$ respectively. We also employed the Euclidean metric $|x|^2=(x^1)^2+\cdots+(x^{\mathcal D})^2$ for $\mathbb R^{\mathcal D}$. In fact, the celebrated Wiener path-integral measure for the time slicing $t_0<t_1<\cdots<t_N$ is;
\begin{equation} \label {wiener1'}
dW(t_N,\cdots,t_1)=\left( \frac{1}{2\pi \Delta t_N} \right)^{\mathcal D/2} ~ e^{-|x_N-x_{N-1}|^2/2\Delta t_N} ~d^{\mathcal D}x_N \times \cdots \times \left( \frac{1}{2\pi \Delta t_1} \right)^{\mathcal D/2} ~ e^{-|x_1-x_0|^2/2\Delta t_1} d^{\mathcal D}x_1.
\end{equation} 

\noindent Based on quantum mechanical purposes, we are enthusiastic about considering the symmetric form of the Wiener measure, i.e., when initial and final points both have been determined. We have:
\begin{equation} \label {wiener1}
\begin{gathered}
dW(t_N,\cdots,t_1)=\left( 4\pi T \right)^{\mathcal D/2}~ \exp \left( |x_F-x_I|^2/4T \right)\\
\times \left( \frac{1}{2\pi \Delta t_{N+1}} \right)^{\mathcal D/2}~ e^{-|x_F-x_{N}|^2/2\Delta t_{N+1}} ~ d^{\mathcal D}x_N ~ \left( \frac{1}{2\pi \Delta t_N} \right)^{\mathcal D/2} ~ e^{-|x_N-x_{N-1}|^2/2\Delta t_N} ~ \times \cdots \\
\times \cdots \times \left( \frac{1}{ 2\pi \Delta t_1 } \right)^{\mathcal D/2} ~e^{-|x_2-x_1|^2/2\Delta t_2} ~ d^{\mathcal D}x_1 ~ \left( \frac{1}{2\pi \Delta t_1}\right)^{\mathcal D/2}~e^{-|x_1-x_I|^2/2\Delta t_1},
\end{gathered}
\end{equation} 

\noindent which explains the probability measure of the stochastic process of the random movement of a particle initiating its Brownian motion from $x_I$ at $t_0=-T$, and terminating it at $t_{N+1}=T$ with approaching $x_F$. We easily compute
\begin{equation}
\int_{\mathbb R^{\mathcal D}} \cdots \int_{\mathbb R^{\mathcal D}} dW(t_N,\cdots,t_1)=1,
\end{equation}

\noindent meaning $dW$ is a well-defined probability measure.

\par Let $\mathbf C$ be the space of all continuous functions on $I$ which have almost everywhere convergent Fourier expansions, and define
\begin{equation} \label{fractal norm}
d_{\kappa} (f,g) = \text{Sup}_{n \in \mathbb Z} \Big| \int_{g_n}^{f_n} e^{\kappa n^2 x^2/2} ~ dx \Big|
\end{equation}

\noindent for $f,g \in \mathbf C$. It is seen that $(\mathbf C, d_\kappa)$ is a well-defined metric space \cite{Varshovi1}. We refer to $d_\kappa$ as the \emph{fractal norm}. Now, assume that $\mathbf C_{\mathcal N}$ consists of functions $f \in \mathbf C$ with $f_n=0$ for $|n|> \mathcal N$. We refer to such functions as $\mathcal N$\emph{-bounded functions}. By definition, $\mathbf C_{\mathcal N}$ is a real vector space with $D_{\mathcal N}=2\mathcal N+1$ dimensions $f_n$ ($-\mathcal N \leq n \leq \mathcal N$). We define the \emph{Lebesgue fractal measure} on $\mathbf C_{\mathcal N}$ by employing the fractal norm (\ref {fractal norm}) on each coordinate $f_n$ as:\footnote{Here we assume that the quantum states on the string would ultimately be fractal on the whole domain $I$ and hence, admit the massless fractal norm. See \cite{Varshovi1} for more details.}
\begin{equation} \label{Fractal norm}
d\ell_{\kappa}(f_n)=e^{\kappa n^2 f_n^2/2}df_n.
\end{equation}

\par To have a well-defined Wiener measure for Brownian motion in the infinite-dimensional Hilbert space of quantum states on a string we have to utilize the fractal norm (\ref{fractal norm}) to control the jumps at extremely high dimensions and transfer the fractal fluctuations to the ultimately far regions of the Hilbert space. Hence, the modified Wiener measure for infinite-dimensions, the so-called \emph{Wiener fractal measure}, is defined based on this viewpoint for $\mathbf C_{\mathcal N}$ by replacing
\begin{equation} \label {x}
\begin{gathered}
x^n = \int_0^{f_n} e^{\kappa n^2 u^2} du~~~\to ~~~ dx^n = e^{\kappa n^2 f_n^2} df_n,\\
~~~~~~~~~~~~~~~~~~d^{\mathcal D}x ~~~ \to ~~~ e^{\sum_n \kappa n^2 f^2_n} d^{\mathcal D} df_n.
\end{gathered}
\end{equation}

\par Consequently, by assuming the definition (\ref{x}) the Wiener fractal measure is:
\begin{equation} \label {wiener3}
\begin{gathered}
dW_{\mathcal N}(t_N,\cdots,t_1)=\left(4\pi T \right)^{D_{\mathcal N}/2} ~\text{exp} \left( \frac{1}{4T} \sum_n \left[ \int_{f_n(-T)}^{f_n(T)} e^{\kappa n^2 x^2/2} dx \right]^2 \right) \\
\times \left( \frac{1}{2\pi \Delta t_{N+1}} \right)^{D_{\mathcal N}/2} ~ \text{exp} \left(\sum_n \Bigg\{ -\frac{1}{2 \Delta t_{N+1}} \left[ \int_{f_n(t_N)}^{f_n(T)} e^{\kappa n^2 x^2/2} dx \right]^2 \Bigg\} \right) \times \\
\prod_{i=1}^N \Bigg\{ \left( \frac{1}{2\pi \Delta t_i } \right)^{D_{\mathcal N}/2} \text{exp} \left(\sum_n \Bigg\{ -\frac{1}{2 \Delta t_i} \left[ \int_{f_n(t_{i-1})}^{f_n(t_i)} e^{\kappa n^2 x^2/2} dx \right]^2 + \frac{1}{2} \kappa n^2 f_n^2(t_i) \Bigg\} \right) ~  d^{D_{\mathcal N}} f_n(t_i) \Bigg\},
\end{gathered}
\end{equation} 

\noindent which substantially describes the Brownian motion of $\mathcal N$-bounded functions by referring to the fractal norm $\ell_\kappa$. Therefore, the $\mathcal N$-bounded functions undergo the Wiener process of fractality while they move stochastically. In fact, the restriction to $\mathbf C_{\mathcal N}$ is a theoretical necessity since integration in infinite-dimensional spaces is inherently ill-defined. Hence, (\ref {wiener3}) is a well-defined probability measure. Nevertheless, since $\mathbf C = \bigcup_{\mathcal N} \mathbf C_{\mathcal N}$,\footnote{We can easily see that the Hilbert space of the quantum states, is, in fact, the topological closure of $\mathbf C$ for the $L^2$ topology. See \cite{Varshovi1} for more details.} then the path-integral will cover the whole regions of $\mathbf C$ as $\mathcal N \to \infty$.

\par For several numbers, say $\mathcal D$, of separate quantum states $f^i$ the Wiener fractal measure would be:
\begin{equation} \label {WienerD_1}
\begin{gathered}
dW_{\mathcal N}(t_N,\cdots,t_1)=\prod_{i=1}^{\mathcal D} dW^i_{\mathcal N}(t_N,\cdots,t_1)\\=\left(4\pi T \right)^{D_{\mathcal N}\mathcal D/2} ~\text{exp} \left( \frac{1}{4T} \sum_{i=1}^{\mathcal D} \sum_n  \left[ \int_{f^i_n(-T)}^{f^i_n(T)} e^{\kappa n^2 x^2/2} dx \right]^2 \right) \\
\times \left( \frac{1}{2\pi \Delta t_{N+1}} \right)^{D_{\mathcal N}\mathcal D /2} ~ \text{exp} \left( \sum_{i=1}^{\mathcal D} \sum_n  \Bigg\{ -\frac{1}{2 \Delta t_{N+1}} \left[ \int_{f^i_n(t_N)}^{f^i_n(T)} e^{\kappa n^2x^2/2} dx \right]^2 \Bigg\} \right) \times \\
\prod_{i=1}^{\mathcal D} \prod_{j=1}^{N} \Bigg\{ \left( \frac{1}{2\pi \Delta t_j } \right)^{D_{\mathcal N}/2} \text{exp} \left(\sum_n \Bigg\{ -\frac{1}{2 \Delta t_j} \left[ \int_{f^i_n(t_{j-1})}^{f^i_n(t_j)} e^{\kappa n^2x^2/2} dx \right]^2 + \frac{1}{2} \kappa n^2 {f_n^{i}}^2(t_j) \Bigg\} \right) ~   d^{D_{\mathcal N}} f^i_n(t_j) \Bigg\}.
\end{gathered}
\end{equation} 

\par It is clear that $dW_{\mathcal N}(t_N,\cdots,t_1)$ is still a well-defined probability measure;
\begin{equation} \label {Wiener 2}
\underbrace{\int_{\mathbb R^{\mathcal D D_{\mathcal N}}} \cdots \int_{\mathbb R^{\mathcal D D_{\mathcal N} }}}_{N-fold} dW_{\mathcal N}(t_N,\cdots,t_1)= 1.
\end{equation}

\par Because of the included non-local terms calculating the Wiener path-integral (\ref {WienerD_1}) is an extremely complex process, unless we employ appropriate approximating assumptions in the exponent of the Wiener fractal measure. Nevertheless, it should be emphasized that the only well-defined and accurate measure that must be utilized for deriving Green's functions of a quantum field theory is the genuine formulation of (\ref {wiener3}). Thus, any approximated version of the Wiener fractal measure (\ref {wiener3}) will lead to some almost accurate formulas for the solutions and the expectation values of the quantum field theory. In practice, we are enthusiastic in a local approxmated version of $dW_{\mathcal N}(t_N,\cdots,t_1)$ which can be extracted after implementing the two following simplifying assumptions:

\par \textbf{Simplification 1)} Discard or neglect the non-local contributions at high amounts of $f^i_n$s. Upon \textbf{Theorem 1} it means to exclude the quantum fields with high amounts of fractality. Indeed, this assumption means that each $U_k \subset \mathbb R^{D_{\mathcal N}\mathcal D}$ in (\ref {PU}) is compact. However, $U_k$s will be replaced by $\mathbb R^{D_{\mathcal N}\mathcal D}$ at the end of the approximation process.

\par \textbf{Simplification 2)} Suppose that $N \to \infty$. Hence: $\Delta t_i=\theta \to 0$, $1\leq i \leq N$.\\

\par Upon the mentioned simplifications we readily find;
\begin{equation} \label {cece}
\begin{gathered}
\Bigg\{ \int_{f_n(t_{j-1})}^{f^i_n(t_j)}  e^{\kappa n^2 x^2/2} dx \Bigg\}/\Delta t_j \to \partial_t {f^i_n } (t_j) e^{\kappa n^2 {f^i_n}^2 (t_j)/2} \\ =\partial_t f^i_n(t_j)
+ \partial_t f^i_n(t_j) \Big\{ \kappa n^2 {f^i_n}^2 (t_j) \Big\} + \frac{1}{2} \partial_t f^i_n(t_j) \Big\{ \kappa n^2 {f^i_n}^2 (t_j) \Big\}^2 + \cdots
\approx \partial_t f^i_n (t_j),
\end{gathered}
\end{equation}

\noindent provided $\kappa f_n^2 \ll1$. Therefore, after implementing \textbf{Simplification 1} and \textbf{Simplification 2} (\ref {WienerD_1}) turns into:
\begin{equation} \label {wiener5}
\begin{gathered}
dW^{\textbf{1},\textbf{2}}_{\mathcal N}(t_N,\cdots,t_1)\\
\approx \text{exp} \left(\sum_{i}^{\mathcal D} \sum_n \Bigg\{ -\frac{1}{2} \left(\partial_t f^i_n(t_j)\right)^2 \Delta t_j + \frac{1}{2} \kappa n^2 {f^i_n}^2(t_j) \Bigg\} \right) \times C_I^F \times \prod_{j=1}^N \prod_{i=1}^{\mathcal D} \left( \left( \frac{1}{\sqrt{2\pi \theta}} \right)^{D_{\mathcal N}} d^{D_{\mathcal N}} f^i_n(t_j) \right) \\
\approx \text{exp} \left({c \frak T} \int_\Sigma \sum_{i=1}^{\mathcal D}\Bigg\{ -\frac{1}{2 c^2} \left(\partial_t f^i (t,\sigma)\right)^2 + \frac{1}{2} \left( \partial_\sigma f^i(t,\sigma ) \right)^2 \Bigg\} d\sigma dt\right) ~\frak D f~~~~~~~~~~~~~~~~~~~~~~~~~~~~~~~~~~~~~~\\
= \text{exp} \left(-\frac{\frak T}{2} \int_\Sigma \sum_{i=1}^{\mathcal D} \Bigg\{ G_{ij}\eta^{a b} \partial_a f^i(\tau,\sigma) \partial_b f^j (\tau,\sigma) \Bigg\} d \tau d\sigma \right) \frak D f, ~~~~~~~~~~~~~~~~~~~~~~~~~~~~~~~~~~~~~~~~~~~~~~~~~~~
\end{gathered}
\end{equation} 

\noindent wherein we have set:
\begin{equation} \label {weiner6}
c=\frac{\ell \sqrt{\kappa}}{2\pi \sqrt{\theta} },~~~~~~~~~~~~\frak T=\frac{\sqrt{\kappa}}{\pi \sqrt{\theta}},~~~~~~~~~~~\Sigma=[-T,T]\times I,
\end{equation}

\noindent while the volume form $\frak D f$ is:
$$\frak D f=C_I^F \left( \frac{1}{2\pi \theta} \right)^{ND_{\mathcal N} \mathcal D/2} \prod_{i=1}^{\mathcal D} \prod_{j=1}^N d^{D_{\mathcal N}} f^i_n(t_j),$$
\noindent for the overall (initial-to-final) normalization factor
$$C_I^F=\left( 4\pi T \right)^{D_\mathcal N \mathcal D/2} ~\text{exp} \left( \sum_{i=1}^{\mathcal D}\sum_n \Big\{ \int_{f_n(-T)}^{f_n(T)} e^{\kappa n^2 x^2/2} dx \Big\}^2 /4T \right).$$

\par Moreover, in the last line of (\ref {wiener5}) use has been made of the standard notations $dx^0=c dt =c \theta$ and $\eta^{a b}=\text{diag}(1,-1)$ as the Lorentz metric in $2$ dimensions. Additionally, we have used $G_{ij}=\text{diag}(1,\cdots,1)$ as the Euclidean metric in $\mathbb R^{\mathcal D}$. To transfer to physical scales the dimension of $f$, i.e. $T^{\frac{1}{2}} $, must be $M^{\frac{1}{2}}T^{-\frac{1}{2}} L$. Let $\beta$ be such an scaling factor of dimension $M^{\frac{1}{2}}T^{-1} L$ and define $X^i=\beta f^i$. Thus, with this redefinition the Wiener fractal measure turns into:
\begin{equation} \label {wiener5'}
\begin{gathered}
dW^{\textbf{1},\textbf{2}}_{\mathcal N}(t_N,\cdots,t_1)\\
= \text{exp} \left(-\frac{\mathcal T}{2} \int_\Sigma \Bigg\{ G_{ij} \eta^{a b} \partial_a X^i(\tau,\sigma) \partial_b X^j(\tau ,\sigma )  \Bigg\} d\tau d\sigma \right) \frak D X,
\end{gathered}
\end{equation} 

\noindent for the tension of the string $\mathcal T =\frak T \beta^{-2}$, while we have simply considered
$$\frak D X =C_I^F \frac{1 }{ \left( 2\pi \theta \beta^2 \right)^{ND_{\mathcal N}\mathcal D/2} } \prod_{i=1}^{\mathcal D} \prod_{j=1}^N d^{D_{\mathcal N}} X^i_n(t_i) $$
\noindent as the \emph{Feynman measure}. In fact, if we replace $\mathcal T$ with $\frac{1}{2\pi \alpha'}$ for the Regge slope $\alpha'$ and assume $T \to \infty$, then the Wiener fractal measure becomes:
\begin{equation} \label {k-g}
dW^{\textbf{1},\textbf{2}}_{\mathcal N}(t_N,\cdots,t_1)= \text{exp} \left( - \frac{1}{4\pi \alpha'} \int_{\Sigma} G_{ij}\eta^{ab} \partial_a X^i \partial_b X^j ~d\tau d\sigma \right) \frak DX,
\end{equation}

\noindent which resembles the Polyakov path-integral measure in Euclidean spacetime. 

\par However, it should be noted that one should not be too delighted by this achievement. We are still having two major defects in our derivation. First, we have made two unreliable simplifications \textbf{Simplification 1} and \textbf{Simplification 2} that discarded all non-local contributions of the theory. Second, we obtained a copy of string theory in Euclidean space which has no physical similarity to the Polyakov formulation, since the spectrum of strings and their quantum mechanical effects are entirely affected by the Lorentzian metric of spacetime $G_{\mu \nu}=\text{diag}(-1,1,\cdots,1)$. In the next section, we will show that string theory is not only the most accurate method to consider the non-local effects of the fluctuations of quantum states in the infinite-dimensional Hilbert space of quantum states but it contains more fundamental interpretations that have never been exposed before.


\par
\section{Derivation: Wiener Fractal Measure and the Polyakov Action}
\setcounter{equation}{0}

\par To consider the non-local contributions in the Wiener fractal measure we have to incorporate the non-local terms of
\begin{equation} \label {Non1}
\sum_{i=1}^{\mathcal D} \left[ \int_{f^i_n(t_{j-1})}^{f^i_n(t_j)} e^{\kappa n^2 x^2/2} dx \right]^2/2\Delta t_j=\sum_{i=1}^{\mathcal D}\frac{1}{2} \left( \partial_t f^i_n(t_j) \right)^2 \Delta t_j +\text{non-local~terms}
\end{equation}

\noindent by means of an augmented local field in the Wiener fractal measure formula. Let us define a new family of time-dependent functions on $I$, denoted by $\frak f(t_j,\sigma): [-T,T] \times I \to \mathbb R$, for $1 \leq j \leq N$, so that the Fourier transform of $\frak f(t_j,\sigma)$, shown by $\frak f_n(t_j)$, is defined by the following equation:
\begin{equation} \label {Non2}
\begin{gathered}
\lim_{\tau \to 0} \sum_{i=1}^{\mathcal D} \left[ \int_{f^i_n(t_j-\tau)}^{f^i_n(t_j)} e^{\kappa n^2 x^2/2} dx \right]/\tau = \sum_{i=1}^{\mathcal D} \frac{d}{dt}\Big|_{t=t_j} \Biggl( \int_0^{f^i_n(t)} e^{\kappa n^2 x^2/2} dx \Biggr)\\
=\Biggl( \sum_{i=1}^{\mathcal D} \frac{d}{dt}\Big|_{t'=t_j}f^i_n(t') \Biggr) e^{\sum_{i=1}^{\mathcal D}\kappa n^2 {f^i_n(t_j)}^2/2}= \left( \sum_{i=1}^{\mathcal D} \frac{df^i_n}{dt}(t_j) \right) + c n \frak f_{-n}(t_j) .
\end{gathered}
\end{equation}

\par Therefore, as the kinetic term of the Lagrangian density $\mathcal L_{K}(\tau,\sigma)$ on the world-sheet the formula of (\ref {Non1}) turns to:
\begin{equation} \label {Non3}
\mathcal L_{K}(\tau,\sigma) = \sum_{i=1}^{\mathcal D}\frac{1}{2} \left( \partial_\tau f^i(\tau,\sigma) \right)^2 + \frac{1}{2} \left( \partial_\sigma \frak f(\tau,\sigma) \right)^2 + \left( \sum_{i=1}^{\mathcal D} \partial_\tau f^i(\tau,\sigma) \right) \partial_\sigma \frak f (\tau,\sigma),
\end{equation}

\noindent where as mentioned above $\tau = c t$ for $c={\ell \sqrt{\kappa}}/{\pi \sqrt{\theta}}$ and the use of (\ref{Derivative}) has been made. Hence, the total Lagrangian density on the world-sheet takes the following form:
\begin{equation} \label {Non4}
\mathcal L (\tau,\sigma) = - \frac{1}{2} \sum_{i=1}^{\mathcal D} \left( \left( \partial_\tau f^i (\tau,\sigma) \right)^2 - \left( \partial_\sigma f^i (\tau,\sigma) \right)^2 \right) - \frac{1}{2} \Bigl( \partial_\sigma \frak f (\tau,\sigma) \Bigr)^2 - \partial_\tau \Bigl( \sum_{i=1}^{\mathcal D} f^i (\tau,\sigma) \Bigr)\partial_\sigma \frak f (\tau,\sigma) .
\end{equation}

\par However, although $\frak f$ includes highly complicated non-local terms it is still a dependent quantum entity with no independent integration in the Wiener fractal measure. Let us introduce an independent quantum field $f^0(t,\sigma): [-T,T]\times I \to \mathbb R$, the so-called \emph{non-locality field}, which plays the role of $\frak f$ in the Lagrangian density but is fixed at $\frak f$ at time sections $t_i$ by inserting appropriate Dirac delta functions. Therefore, the Lagrangian density (\ref {Non4}) would be of the following form
\begin{equation} \label {Non5}
\mathcal L (\tau,\sigma) = - \frac{1}{2} \sum_{i=1}^{\mathcal D} \left( \left( \partial_\tau f^i (\tau,\sigma) \right)^2 - \left( \partial_\sigma f^i (\tau,\sigma) \right)^2 \right) - \frac{1}{2} \Bigl( \partial_\sigma f^0 (\tau,\sigma) \Bigr)^2 - \partial_\tau \Bigl( \sum_{i=1}^{\mathcal D} f^i (\tau,\sigma) \Bigr)\partial_\sigma f^0 (\tau,\sigma),
\end{equation}

\noindent while the Wiener fractal measure will be accompanied by multiple Dirac delta functions and the corresponding overall integration volume forms for the non-locality field $f^0$. Working in the Fourier space the Wiener fractal measure will contain additional terms as below:
\begin{equation} \label {Non6}
\prod_{j=1}^N\left( \frac{1}{2\pi \Delta t_j} \right)^{D_{\mathcal N}/2} \exp \left(\sum_n  -\frac{1}{2 \Delta t_j} \Bigl( f^0_n(t_j) - \frak f_n(t_j) \Bigr)^2 \right)~d^{D_{\mathcal N}} f^0_n(t_j).
\end{equation}

\noindent Strictly speaking, the Wiener fractal measure for $\mathcal D = D-1$ dimensions turns to:
\begin{equation} \label {Non66}
\begin{gathered}
dW_{\mathcal N}(t_N,\cdots,t_1) \\
 = \frak C_N ~ \text{exp} \left(  - \frac{1}{2} \sum_{j=1}^N \sum_n \Biggl[ \sum_{i=1}^{D-1}  \Biggl( \Bigl( \partial_t f_n^i (t_j) \Bigr)^2 \Delta t_j - \Bigl( n f^i_n (t_j) \Bigr)^2 \Biggr) + \Bigl( n f^0_n (t_j)  \Bigr)^2 + \frac{1}{ \Delta t_j} \Bigl( f^0_n(t_j) - \frak f_n(t_j) \Bigr)^2 \Biggr] \right) \\
\times ~ \text{exp} \left(  - \sum_{j=1}^N \sum_n \Biggl[ \Biggl( \sum_{i=1}^{D-1}   \Bigl( \partial_t f_n^i (t_j) \Bigr) \Delta t_j \Biggr) c nf_{-n}^0(t_j) \Biggr] \right) \times \prod_{j=1}^N \Biggl[~\left( \prod_{i=1}^{D-1} d^{D_{\mathcal N}} f^i_n(t_j) \right) ~d^{D_{\mathcal N}} f^0_n(t_j) \Biggr],~~~~~~~~~~~~
\end{gathered}
\end{equation} 

\noindent for
\begin{equation} \label{Non66-1}
\begin{gathered}
\frak C_N =\left(4\pi T \right)^{D_{\mathcal N}(D-1)/2} ~\text{exp} \left( \frac{1}{4T} \sum_{i=1}^{D-1} \sum_n  \left[ \int_{f^i_n(-T)}^{f^i_n(T)} e^{\kappa n^2 x^2/2} dx \right]^2 \right) \\
\times \left( \frac{1}{2\pi \Delta t_{N+1}} \right)^{D_{\mathcal N} (D-1) /2} ~ \text{exp} \left( \sum_{i=1}^{D-1} \sum_n  \Bigg\{ -\frac{1}{2 \Delta t_{N+1}} \left[ \int_{f^i_n(t_N)}^{f^i_n(T)} e^{\kappa n^2x^2/2} dx \right]^2 \Bigg\} \right) \times \prod_{j=1}^{N}  \left( \frac{1}{2\pi \Delta t_j } \right)^{D_{\mathcal N} D/2}.
\end{gathered}
\end{equation}

\noindent Obviously, the above replacements are accurate as $\Delta t_j \to 0$ for all $1\leq j \leq N$. At this level of precision we can make our first approximation due:\\

\par \textbf{Approximation I:} \emph{According to the consecutive Dirac delta functions incorporated into the Wiener fractal measure each $\frak f_n(t_j)$ can be replaced by the two first terms of the following formula}
\begin{equation} \label {Non7}
\frak f_n(t_j)=f^0_n(t_{j-1}) + \Delta t_j \partial_t f_n^0(t'^n_{j-1}) + \varepsilon (t_{j-1}),
\end{equation}

\noindent \emph{where in the context of the mean value theorem, $t'^n_{j-1} \in (t_{j-1}, t_j)$ represents an appropriate time at which the derivative equals the mean slope of the function, while $\varepsilon(t_{j-1})$ accounts for the contribution of the non-differentiability of the Fourier modes $f^0_n(t)$.}\\

\par According to \textbf{Approximation I} the contribution of the Dirac delta functions in the Fourier space formulation of the Lagrangian ${L}^{Dirac}$ would be:
\begin{equation} \label {Non8}
\begin{gathered}
{L}^{Dirac} (t_j)= -\sum_{n} \frac{1}{2\Delta t_j} \Bigl( f^0_n(t_j) - f^0_n(t_{j-1}) - \Delta t_j \partial_t f^0_n(t'^n_{j-1}) \Bigr)^2 ~~~~~~~~~~~~~~~~~~~~~~~~~~~~~~\\
= -\sum_{n} \frac{1}{2\Delta t_j} \Bigl( f^0_n(t_j) - f^0_n(t_{j-1}) \Bigr)^2 -\sum_{n} \frac{1}{2\Delta t_j} \Delta t_j^2 \left( \partial_t f^0_n(t'^n_{j-1}) \right)^2 \\
+ \sum_{n} \frac{1}{\Delta t_j} \Bigl( f^0_n(t_j) - f^0_n(t_{j-1}) \Bigr) \Delta t_j \partial_t f^0_n(t'^n_{j-1}),~~~~~~~~~~~~~~~~~~~~~~~~~
\end{gathered}
\end{equation}

\noindent which in the limit of $\Delta t_j \to 0$ becomes:
\begin{equation} \label {Non9}
{L}^{Dirac} (t_j) =- \frac{\Delta t_j}{2} \Big\{ \sum_{n}  \Bigl( \partial_tf^0_n(t_j) \Bigr)^2 - 2 \sum_{n} \partial_t f^0_n(t_j) \partial_t f^0_n(t'^n_{j-1}) \Big\} + L^{irr}(t_j,t_{j-1}),
\end{equation}

\noindent where
\begin{equation} \label {irr}
L^{irr}(t_j,t_{j-1})= \frac{\Delta t_j}{2} \sum_{n}  \Bigl( \partial_tf^0_n(t'^n_{j-1}) \Bigr)^2,
\end{equation}

\noindent the so-called \emph{irrelevant Lagrangian}, has no contributions in the essential time sections of the Wiener fractal path-integral measure due to $df^0_n(t_j)$s. Strictly speaking, the fundamental structure of the functional measure in the Wiener fractal measure is based on the Fourier mode variables $f_n(t_j)$, which are inherently irrelevant to the variables involved in $L^{\text{irr}}(t_j,t_{j-1})$. In addition, we must emphasize that there is no well-defined way to extend our path-integral formalism to the intermediate time sections due to $t'^n_j$s included in $L^{\text{irr}}(t_j,t_{j-1})$.In essence, $t'^n_j$ does not actually provide a well-defined time section for several reasons: \textbf{a)} $t'^n_j$ depends on $n$, and hence each Fourier mode $f^0_n$ has its own intermediate time $t'^n_j$; \textbf{b)} $t'^n_j$ depends on the analytic structure of $f^0_n(t)$, and consequently, the path-integration over $f^0_n(t)$ will affect the value of $t'^n_j$; and \textbf{c)} there may be more than one $t'^n_j$ in $(t_{j-1}, t_j)$ that satisfies the conditions of the mean value theorem, and there is no specific criterion to choose one of them.

\par Upon the above arguments, the irrelevant Lagrangian $L^{irr}(t_j,t_{j-1})$, from an optimistic viewpoint, suggests a separate free theory with no well-defined Lagrangian density and path-integral formualtion. Thus, the non-local term $\sum_n \partial_t f^0_n(t_j) \partial_t f^0_n(t'^n_{j-1})$ within $L^{Dirac}(t_j)$ does not, in fact, constitute a well-defined interaction term. All in all, the irrelevant Lagrangian $L^{\text{irr}}(t_j,t_{j-1})$ lacks both a well-defined path-integral basis and any relevant physical impact on observable outcomes derived from the final Wiener fractal measure formulation. Therefore, it should be either excluded from $L^{Dirac}(t_j)$ or treated as an overall factor for the Wiener fractal measure (\ref{Non66}), hidden in $\mathfrak{C}_N$. Consequently, the non-local term $\sum_n \partial_t f^0_n(t_j) \partial_t f^0_n(t'^n_{j-1})$ in $L^{Dirac}(t_j)$ necessitates approximation by the closest well-defined local term. Hence, our next step involves implementing an appropriate approximation strategy.\\

\par \textbf{Approximation II:} \emph{The non-local term $\sum_n\partial_t f^0_n(t_j) \partial_t f^0_n(t'^n_{j-1})$ in ${L}^{Dirac} (t_j)$ is replaced by its closest local term $\sum_n \left( \partial_tf^0_n(t_j) \right)^2$.}\\

\par Hence, according to \textbf{Approximation II} the augmented term ${L}^{Dirac} (t_j)$ would be: 
\begin{equation} \label {Non9'}
{L}^{Dirac} (t_j) = + \frac{\Delta t_j}{2} \sum_{n}  \Bigl( \partial_tf^0_n(t_j) \Bigr)^2.
\end{equation}

\noindent Consequently, in the limit of $\Delta t_j \to 0$ and with the hypothesis of \textbf{Approximation I}, after performing the localization mechanism proposed in \textbf{Approximation II} the Lagrangian density (\ref {Non5}) becomes:
\begin{equation} \label {Non10}
\begin{gathered}
\mathcal L (\tau,\sigma) = - \frac{1}{2} \sum_{i=1}^{D-1} \left( \left( \partial_\tau f^i \right)^2 - \left( \partial_\sigma f^i \right)^2 \right) + \frac{1}{2} \left( \left( \partial_\tau f^0 \right)^2 - \left( \partial_\sigma f^0 \right)^2 \right) + \mathcal L_{ext}(f^0,f^i) \\
= - \frac{1}{2} G_{\mu \nu} \eta^{ab} \partial_a f^\mu \partial_b f^\nu + \mathcal L_{ext}(f^0,f^i),~~~~~~~~~~~~~~~~~~~~~~~~~~~~~~~~~~~
\end{gathered}
\end{equation}

\noindent where we used the Lorentzian metrics $\eta_{ab} = \text{diag} (-1,1)$ and $G_{\mu \nu}= \text{diag} (-1, 1, \cdots, 1)$ for $a,b \in \{0,1\}$ and $\mu, \nu \in \{ 0, 1, \cdots, D-1 \}$, while we have defined the extra term $\mathcal L_{ext}(f^0,f^i)$ as
\begin{equation}
 \mathcal L_{ext}(f^0,f^i) = - \partial_\tau \Bigl( \sum_{i=1}^{\mathcal D} f^i (\tau,\sigma) \Bigr)\partial_\sigma f^0 (\tau,\sigma).
\end{equation}

\par Now we are ready to focus on $\mathcal L_{ext}(f^0,f^i)$ as the asymmetric term of the Lagrangian density $\mathcal L (\tau,\sigma)$. Although its emergence is awkward because it breaks the crucial symmetric structures like the general coordinate transformation on the world-sheet and the Lorentz invariance of the target space, it can be seen that $\mathcal L_{ext}(f^0,f^i)$ is redundant and must be eliminated from the Lagrangian density of the Wiener fractal measure to provide a free theory. Indeed, according to the substantial way of introducing the non-locality field into the Wiener fractal measure, the ultimate formula must be inherently symmetric for the transformation $f^0_n(t_j) \leftrightarrow -f^0_n(t_j)$.\footnote{In other words, we can equivalently avoid replacing $\frak f_n(t_j)$ with $f_n^0(t_j)$ inside the genuine form of the Lagrangian density (\ref {Non4}) (before writing the Lagrangian density (\ref{Non5})) and take the integrations of $f^0_n(t_j)$s for the corresponding Dirac delta functions (\ref {Non6}) separately.} In this regard, since the genuine formula is invariant with respect to this transformation the final formulation of the Wiener fractal measure must be invariant too. Therefore, the total Lagrangian density $\mathcal L (\tau,\sigma) $ must be an even functional of $f^0_n(t_j)$ and its time derivative, which means that $\mathcal L_{ext}(f^0,f^i)$ has appeared because of our approximations and must be removed from the ultimate formulation of the Wiener fractal measure.

\par Meanwhile, we can demonstrate the above statement by various reasonings including the symmetriy arguments and direct calculations. One of the symmetry arguments stems from the inherent time-reversal symmetry of the genuine Wiener fractal measure (\ref{WienerD_1}) as we started from (\ref{wiener1}). More precisely, the Wiener fractal measure (\ref{WienerD_1}) is symmetric with respect to $t \leftrightarrow -t$ and hence, the Lagrangian density $\mathcal L (\tau,\sigma) $ must be time-reversal too. On the other hand, $\mathcal L_{ext}(f^0,f^i)$ manifestly breaks this symmetry and hence, must be discarded from the final formulation of the Wiener fractal measure.

\par The other symmetry argument stems from the invariance of the Wiener fractal measure with respect to the transformation $\sigma \leftrightarrow \ell -\sigma$. It means that the ultimate formulation of Brownian motion of quantum states on a string would not be affected if we flip the string. Therefore, the Lagrangian density $\mathcal L (\tau,\sigma) $ must be symmetric with respect to $\sigma \leftrightarrow \ell -\sigma$. However, $\mathcal L_{ext}(f^0,f^i)$ violates this symmetry and consequently, it must be eliminated from the final formula of the Wiener fractal measure.

\par The fourth proof for the claim that $\mathcal L_{ext}(f^0,f^i)$ must be removed from the Lagrangian density $\mathcal L (\tau,\sigma) $ is based on the direct calculations. Essentially, the contributions of $\mathcal L_{ext}(f^0,f^i)$ into the Wiener fractal path-integral come from the integrations of the following form
\begin{equation} \label {GGG}
\int_{f^0_n(t_j)\in \mathbb R} \Biggl[ \int_{f^i_n(t_j) \in U} G\Bigl( f^i_n(t_j), f^0_n(t_j) \Bigr) e^{ \alpha_n f^i_n(t_j) f^0_n(t_j) } ~ df^i_n(t_j) \Biggr]~df^0_n(t_j)
\end{equation}

\noindent for some functional $G$ depending on even powers of $f^i_n(t_j)$ and $f^0_n(t_j)$, and for $\alpha_n$ an overall constant. Employing the Taylor expansion we readily find that (\ref {GGG}) would be of the following form
\begin{equation} \label {GGGG}
\begin{gathered}
\int \Bigg\{ \int G\Bigl( f^i_n(t_j), f^0_n(t_j) \Bigr) \Bigl( \sum_{m=0}^\infty \frac{\alpha_n^m }{m!} \left( f^i_n(t_j) f^0_n(t_j)  \right)^m \Bigr) ~ df^i_n(t_j) \Bigg\}~df^0_n(t_j)\\
=\int \Bigg\{ \int G\Bigl( f^i_n(t_j), f^0_n(t_j) \Bigr) \Bigl( \sum_{m=0}^\infty \frac{ \alpha_n^{2m} }{(2m)!} \left( f^i_n(t_j) f^0_n(t_j)  \right)^{2m} \Bigr) ~ df^i_n(t_j) \Bigg\}~df^0_n(t_j) \\
= \int \Bigg\{ \int G\Bigl( f^i_n(t_j), f^0_n(t_j) \Bigr) e^{\alpha^2_n/2 \left( f^i_n(t_j) f^0_n(t_j) \right)^2 +~ \cdots } ~ df^i_n(t_j) \Bigg\}~df^0_n(t_j).~~~~~~~~~~~
\end{gathered}
\end{equation}

\noindent As we see from (\ref {GGGG}) the contribution of $\mathcal L_{ext}(f^0,f^i)$ in the total Lagrangian density $\mathcal L (\tau,\sigma) $ starts from a monomial of order four, meaning $\mathcal L_{ext}(f^0,f^i)$ is inherently an interaction term and must be removed from the kinetic part of the Lagrangian density as a free quantum theory. 

\par In fact, discarding $\mathcal L_{ext}(f^0,f^i)$ from the ultimate formulation of the Wiener fractal measure can be considered as the third approximation method. However, it seems that this elimination tries to turn back the Lagrangian density $\mathcal L (\tau,\sigma) $ as close as possible to its more original form. So it may be considered as an improvement procedure. Therefore, we just do it without attributing statements like approximation or modification. All in all, the ultimate and the most accurate formula of the total Lagrangian density $\mathcal L (\tau,\sigma)$ which respects the whole fundamental symmetries of the free theory is:
\begin{equation}
\begin{gathered}
\mathcal L (\tau,\sigma) = - \frac{1}{2} G_{\mu \nu} \eta^{ab} \partial_a f^\mu \partial_b f^\nu.
\end{gathered}
\end{equation} 

\par Now, with employing the parameters and definitions we used in (\ref{wiener5}), (\ref{weiner6}), and (\ref{wiener5'}), and after applying \textbf{Approximation I} and \textbf{Approximation II} we will obtain the final version of the approximated Wiener fractal measure $dW_{\mathcal N}^{\textbf{I},\textbf{II}}$ as below:
\begin{equation} \label {polyakov1}
dW_{\mathcal N}^{\textbf{I},\textbf{II} } (t_N,\cdots,t_1)= \text{exp} \left( - \frac{1}{4\pi \alpha'} \int_{\Sigma} G_{\mu \nu}\eta^{ab} \partial_a X^\mu \partial_b X^\nu ~d\tau d\sigma \right) \frak DX,
\end{equation}

\noindent for
\begin{equation} \label{String1}
\frak D X =C_I^F \frac{1 }{ \left( 2\pi \theta \gamma^2 \right)^{ND_{\mathcal N} D/2} } \prod_{\mu=0}^{ D - 1 } \prod_{j=1}^N d^{D_{\mathcal N}} X^\mu_n(t_i).
\end{equation}

\par Any coordinate transformation on the world-sheet $\Sigma$ may cause a change in the Lorentzian metric $\eta_{ab}$ and the corresponding volume form $d\tau d\sigma$ respectively to $\gamma_{ab}$ and $\sqrt{|\gamma|} d^2\sigma$ for $\gamma = \det{(\gamma_{ab})}$. Thus, a more general formulation of the approximated Wiener fractal measure (\ref{polyakov1}) is:
\begin{equation} \label {polyakov2}
dW_{\mathcal N}^{\textbf{I},\textbf{II}} (t_N,\cdots,t_1)= \text{exp} \left( - \frac{1}{4\pi \alpha'} \int_{\Sigma} \sqrt{|\gamma|} ~G_{\mu \nu}\gamma^{ab} \partial_a X^\mu \partial_b X^\nu ~d^2\sigma \right) \frak DX,
\end{equation}

\noindent which is the precise formulation of the genuine Polyakov path-integral measure whenever we consider $\mathcal N \to \infty$ in our calculations.

\par As seen above, the Polyakov path-integral measure $dW_{\mathcal N}^{\textbf{I},\textbf{II}}$ was extracted with the least amount of approximation from the Wiener fractal measure. Consequently, the well-known $\textbf{Diff} \times \textbf{Weyl}$ invariance of the action is, in fact, the genuine inherent symmetry of the Wiener fractal measure (\ref {Non66}) at the highest level of accuracy. Therefore, it is naturally expected that any interaction term within the theory must respect such intrinsic geometric structures. This point of view readily leads us to the standard interpretation of the Polyakov path-integral in which the interactions of strings are formulated as the gluing, the cutting, and the sewing processes of the world-sheets of strings given in terms of a unique (topological) coupling constant based on the Euler characteristic number.

\par It is obvious that the Brownian motion of the quantum states on strings has nothing to do with the coordinate transformations on the world-sheet and hence, must be $\textbf{Diff}$-invariant. In this way, the only interaction term that can be added to the power of the Wiener fractal measure next to the Polyakov Lagrangian density is the scalar curvature $R$ modulo some constant which only causes the action to be shifted by a constant provided world-sheets have two dimensions, i.e.:
\begin{equation} \label {Pol-Action}
dW_{\mathcal N}^{\textbf{I},\textbf{II}} \propto e^{S_{\textbf{Pol}} + S_{\textbf{Eul}} } ~ \frak DX,
\end{equation}

\noindent for the Polyakov action
\begin{equation} \label {Pol-Action2}
S_{\textbf{Pol}}=-\frac{1}{4\pi \alpha'} \int_{\Sigma} \sqrt{|\tilde{\gamma}|} ~G_{\mu \nu}\tilde{\gamma}^{ab} \partial_a X^\mu \partial_b X^\nu ~ d^2 \sigma
\end{equation}

\noindent with the Euclidean metric $\tilde{\gamma}_{ab}$,\footnote{Transferring to the Euclidean metric is based on the same footing we have in the mathematics of string theory.} and the Euler action
\begin{equation} \label {Pol-Action3}
S_{\textbf{Eul}}= - \lambda \mbox{\Large$\chi$}_{\Sigma} = - \lambda \int_{\Sigma} \sqrt{|\tilde{\gamma}|} ~R ~d^2 \sigma = - \lambda \Bigl( N_2(\Sigma) - N_1(\Sigma) + N_0 (\Sigma) \Bigr),
\end{equation}

\noindent wherein $\mbox{\Large$\chi$}_{\Sigma}$ is the Euler characteristic of the world-sheet $\Sigma$, while the terms $N_2(\Sigma)$, $N_1(\Sigma)$, and $N_0(\Sigma)$ represent the number of 2-cells (surfaces), 1-cells (boundaries), and 0-cells (corners) in $\Sigma$, respectively \cite{GSW, Polchinski}.

\par As we argued above the $\textbf{Diff}$-invariance of the Brownian motion of quantum states stems from the physics of the stochastic process. However, the $\textbf{Weyl}$-invariance emerges automatically from the very formulation of the Wiener path-integral (\ref{polyakov2}). Therefore, the invariance of $S_{\textbf{Pol}}+S_{\textbf{Eul}}$ with respect to general coordinate transformations on the world-sheet $\Sigma$ and the Weyl transformation of the metric $\tilde{\gamma}_{ab}$ causes an oversummation on equivalent geometries for each fixed topology. Hence, as a theory of Brownian motion of quantum states the Wiener fractal measure (\ref{polyakov2}) must be computed in a fixed gauge or equivalently divided by $\mathcal V_{\textbf{Diff} \times \textbf{Weyl} }$, the volume of the Lie group $\textbf{Diff}(\Sigma) \times \textbf{Weyl}(\tilde{\gamma})$, for extracting the observable (gauge invariant) results and amplitudes. For instance, the corresponding $\textbf{S}$-matrix elements due to the Brownian motion of quantum states on strings must be given by the well-known formula of string theory \cite{Polchinski}:
\begin{equation} \label {Pol-Action4}
\textbf{S}_{j_1 \cdots j_n}(k_1,\cdots,k_n)= \frac{1}{\mathcal V_{\textbf{Diff} \times \textbf{Weyl} }} \sum_{\text{Compact~Topologies}} \int dW^{\textbf{I},\textbf{II} } \frak D\tilde{\gamma} \Biggl( \prod_{i=1}^n \int_{\Sigma} \sqrt{\tilde{\gamma}(\sigma_i)} ~ V_{j_i}(k_i\sigma_i) ~ d^2 \sigma_i \Biggr),
\end{equation}

\noindent wherein, here, $dW^{\textbf{I},\textbf{II} } = \lim_{\mathcal N \to \infty}dW_{\mathcal N}^{\textbf{I},\textbf{II} }$ is only calculated for equivalent topologies, $\frak D\tilde{\gamma}$ provides the summation over all $\textbf{Diff}(\Sigma) \times \textbf{Weyl}(\tilde{\gamma})$ transformations, and $V_{j}(k_j,\sigma)$ is the vertex operator for the internal state $j$ and $D$-momentum $k^\mu$. It must be clarified that the summation over \emph{compact topologies} and also the equivalence relation between quantum states on strings and the vertex operators $V_{j}(k_j,\sigma)$ follow similarly from the standard argumentations in the mathematics of string theory \cite{Polchinski}.

\par It has already been shown in \cite{Varshovi1, Varshovi3} that the Brownian motion of quantum states on any compact space manifold leads to Feynman's path-integral formulation of the corresponding Klein-Gordon or Yang-Mills theory provided the non-locality field is neglected in low energy scenarios.\footnote{Indeed, in \cite{Varshovi2} and \cite{Varshovi3}, we have also shown that studying the Brownian motion of scalar and gauge quantum states will automatically produce the Einstein-Hilbert action in the genuine formula of the Wiener fractal measure whenever the space metric evolves via the entropic Hamilton-Perelman Ricci flow. Additionally, it has been proven in \cite{Varshovi3} that to compensate for the non-local effects emerging in the Brownian motion of the gauge quantum states, the spinor matter fields arise spontaneously in the Wiener fractal measure leading to the well-known formulation of the Standard Model. All in all, the achievements of \cite{Varshovi1, Varshovi2, Varshovi3} and what we have extracted here for strings demonstrate that the study of the Brownian motion of quantum states on a compact base manifold will disclose basic information about the most fundamental corresponding quantum mechanical effects.} However, here, the strategy is developed for the highest level of accuracy which comprises the high energy contributions of the non-local effects in the Wiener fractal measure formula. Therefore, the Polyakov action is inherently a theoretical development of the Brownian motion of quantum states and hence, of the existing framework of scalar quantum field theory in which non-local effects have been included in the framework of path-integral formulation with the utmost accuracy.

\par The above achievements substantially demonstrate that the Polyakov action is not a mysterious formulation uniquely describing the fundamental forces of nature, but rather a standard formulation of the Brownian motion of quantum states on any arbitrary one-dimensional microscopic objects like strings. These arbitrary one-dimensional objects physically include any three-dimensional object whose one dimension is extremely larger than the other two dimensions. Therefore, the Polyakov path-integral formulation also provides the only possible mathematical framework for investigating the fluctuations of quantum states in brain neurons or in any stringy subunits of neurons. This point brings one of the most delicate scientific topics, the quantum mind, into the realm of string theory studies, since, in fact, the quantum observer consciousness and its interaction with quantum phenomena is one of the most fundamental and at the same time most incomprehensible issues of quantum mechanics. We will address this issue in more detail in the next sections.


\par
\section{Reconsideration: String Theory and Unreliable Interpretations}
\setcounter{equation}{0}

\par According to the above achievements, a quantum theory of the Brownian motion of the quantum states of a string will exactly match the bosonic string theory formulations. The inquiry into whether the Polyakov action is coincidentally consistent with Wiener's formulation of these Brownian motions, or whether this consistency is rooted in deeper concepts in the foundations of quantum mechanics, leads us to more serious challenges in our interpretations of quantum mechanics and string theory. However, our distinct interpretation of $X^0$ as the field of non-locality shows that apparently, our interpretation of other $X^\mu$ fields also has nothing to do with the dimensions of spacetime. Strictly speaking, there is no guarantee that the $X^\mu$ bosonic fields in the Polyakov action are actually related to the string quantum mechanical fluctuations in spacetime dimensions, since at least the interpretation of the non-locality field $X^0$ does not correspond to its prevalent interpretation in string theory.

\par Therefore, it appears that the discovery of the Polyakov action in the study of Brownian motion of quantum states on strings will provide an entirely new and unpredictable interpretation of the achievements of the bosonic string theory. Also, this conclusion can be readily extended to the elegant results of superstring theory. Here, the most basic question for our perusal is whether string theory is an accurate quantum model for studying the Brownian motions in the Hilbert space of quantum states on strings or a complicated quantum theory to describe the fundamental forces of nature. The term "string" encompasses any arbitrary thread, including microscopic brain neurons or their stringy subunits. Therefore, more precisely, the above query can be raised with a more pragmatic insight: Is string theory a successful quantum model for describing quantum interactions of brain neurons or a physical theory for understanding the fundamental interactions of nature?

\par Several signs may lead us to suspect that we misinterpret string theory and its achievements. The impressive achievements of string theory, despite their many similarities to mathematical formulations and the quantum effects of the fundamental forces of nature, have never been able to provide a more reliable and powerful phenomenological replacement for the Standard Model and beyond. In fact, the most significant flaw of string theory is that it has not yet produced a rigorous description of the observed phenomena in particle physics. While (super)string theory aims to provide a beautiful unified framework that includes gravity and incorporates quantum mechanics, it still remains largely theoretical and lacks experimental verification.\footnote{See \cite{Smolin, Woit, Dawid, Ritson, Rovelli, Rovelli1998, Rovelli2003, Rovelli2016, Conlon, Dine, Elis, Glashow, Marchesano2024, Smolin2001, Weinberg, BBS, Greene} for more useful discussions.} Indeed, over recent decades, a central query in string phenomenology has been understanding the relationship between the Standard Model of particle physics and string theory.

\par String theory initially heralded as a promising candidate for a unified theory of everything, has captivated physicists with its beautiful, elegant mathematical structure and its potential to reconcile quantum mechanics with general relativity. The pursuit of producing the Standard Model has led to the development of various superstring theory construction strategies that aim to reproduce the matter content, gauge groups, and other fundamental aspects of the model via sophisticated superstring mathematical tools and procedures \cite{GSW2, BBS, Polchinski2}. However, despite significant progress and a much deeper understanding of the mathematics underlying string compactifications,\footnote{See \cite{Polchinski2, Greene, Freedman, Douglas, Bars, Candelas, Kachru} for more details.} constructing fully realistic string vacua that incorporate all observed features such as the Yukawa couplings,\footnote{See \cite{Blesneag, Font, Gepner, Candelas1987, Sotkov} and the references therein for more discussions.} the Higgs sector,\footnote{See \cite{Abel2021, Lim} as two nice presentations of the issue. See also \cite{Barth, Jenni} for more related investigations.} supersymmetry breaking,\footnote{See \cite{Douglas2007, DSW, Dixon, Freedman2012, Murthy, Kounnas, Mayr}.} the detailed structure of the Standard Model parameters,\footnote{See \cite{Marchesano2024, Kane, Bailin, Cvetic, Ibanez} and the references therein.} and the observed cosmological constant\footnote{See \cite{Bousso, Berglund2023, Witten2001, Kachru, Susskind2003, Kachru2003', Douglas2007, DeWolfe} and the references therein for more discussions about the cosmological constant in string theory.} remains an immense challenge.\footnote{For more discussions about the aims, achievements, and far-reaching dreams of string theory see \cite{Dine, Weinberg1992, Green1986, Duff1998}.}

\par However, despite its theoretical allure, string theory is, in fact, fraught with significant defects that challenge its validity and efficacy as a comprehensive physical theory. Let us have a brief survey about the disabilities and shortcomings of string theory in describing experimental quantum physics and the corresponding laboratory observations. As specified above one of the most glaring defects of string theory is its lack of empirical verification.\footnote{See for example \cite{Smolin, Woit, Dawid, Ritson, Rovelli, Rovelli1998, Rovelli2003, Rovelli2016, Conlon}.} Unlike quantum mechanics and general relativity, which have withstood rigorous experimental testing, string theory remains speculative. The energy scales required to test predictions of string theory are far beyond the reach of current technology, making direct experimental validation virtually impossible. This lack of empirical evidence has led to criticism that string theory is more of a mathematical curiosity than a physical theory.\footnote{See for example \cite{Rovelli2016, Conlon, Smolin}.} Moreover, string theory's predictive power is another point of contention. The theory encompasses an enormous landscape of possible solutions, known as the \emph{string theory landscape}, which contains around $10^{500}$ different vacua.\footnote{See \cite{Read, Candelas, Douglas2019, Bena, Douglas2003, Ashok, Taylor} and the references therein for more more details.} This vast number of potential solutions makes it exceedingly difficult to derive unique, testable predictions. Instead of providing precise answers, string theory often predicts a multitude of possible outcomes, reducing its utility in making definitive scientific predictions.\footnote{See for example \cite{Dawid2007, Dawid, Smolin, Kumar}.}

\par The mathematical complexity of string theory is both a superb strength and a fatal weakness. While the sophisticated mathematics behind string theory can elegantly describe high-dimensional spaces and various physical phenomena, it also renders the theory highly abstract and difficult to interpret physically.\footnote{See \cite{Smolin, Woit}.} The necessity to work in higher dimensions (typically $26$, $10$, or $11$ for $M$-theory) further complicates the theory, making it less accessible and more prone to ambiguities.\footnote{See \cite{Dawid, Dawid2007, Smolin, Woit}.} Furthermore, string theory is inherently background-dependent, meaning it requires a pre-defined spacetime to describe the dynamics of strings. This starkly contrasts general relativity, which is a background-independent theory where spacetime itself is dynamic and shaped by the presence of matter and energy. The reliance on a fixed background undermines the fundamental principle that spacetime geometry should emerge from the dynamics of the theory itself.\footnote{See \cite{Smolin, Dawid, Rovelli, Rovelli2003, Rovelli1998} for more related discussions.}

\par The next flaw of string theory is the issue of unification and exclusivity. More precisely, string theory aims to be a \emph{theory of everything}, yet it has struggled to unify all fundamental forces and particles uniquely. The existence of multiple string theories (including \textbf{Type I}, \textbf{Type IIA}, \textbf{Type IIB}, heterotic $SO(32)$, and heterotic $E_8\times E_8$) necessitated the development of $M$-theory, which purportedly unifies these into a single framework \cite{Witten1995, Witten1996}. However, this beautiful unification is more a patchwork of different mathematical theories rather than a seamless, exclusive theory of all physical phenomena.\footnote{See \cite{Dawid, Dawid2007, Smolin, Woit, Rovelli2016} for more discussions.} Hence, it seems that string theory is more mathematics and less physics. 

\par One of the most related and significant theoretical failings of string theory is its incorrect prediction regarding the sign of the cosmological constant. Observations show that the cosmological constant is positive, driving the accelerated expansion of the universe. However, early models of string theory tended to predict a negative or zero cosmological constant.\footnote{See \cite{Carroll, Weinberg1, Weinberg2} and the references therein for more details.} This discrepancy indicates a fundamental issue within the theory's framework in accurately describing the large-scale structure and dynamics of our universe. This fatal flaw of string theory is in fundamental contradiction with its unique role in providing a complete physical model for all the fundamental forces of nature. Is it plausible that string theory, which is at the utmost perfection and beauty of symmetric structures and the relevant sophisticated mathematical aspects,\footnote{See \cite{Gross, Penrose} for two beautiful presentations of the role of symmetries and the corresponding background mathematics in building physical theories.} is talking about other subjects and objects and not the fundamental building blocks of nature and the real world outside?

\par Although mentioned implicitly, we must emphasize that the most notable defect of string theory is its wrong prediction regarding the dimensionality of spacetime. String theory inherently predicts the existence of $26$, $10$, or $11$ spacetime dimensions, depending on the specific mathematical formulation of the theory. It seems that the difference between such numbers and the four dimensions of spacetime is not a problem that can be easily solved in the mathematical framework of the theory and any attempt to untangle this quandary will face us with more complications in our interpretations of string theory.\footnote{See \cite{Smolin, Dawid2007, Woit, Rovelli2016} for more relevant arguments.} While string theorists propose mechanisms such as compactification to reconcile this difference, these mechanisms remain speculative and unproven, adding another layer of complexity and doubt about the theory's physical relevance. With such a prospect, the world seems to be more geometric than physical through the lens of string theory.\footnote{For more similar discussions see \cite{Dawid2007, Woit}.}

\par The unsolvable problem of extra dimensions of spacetime, apart from the mentioned theoretical and experimental ambiguities, makes our interpretations face serious defects and obscurities. More precisely, the other problem of string theory regarding the wrong prediction of spacetime dimensions concerns the philosophical and interpretative challenges. Strictly speaking, beyond the technical and empirical issues, string theory also faces philosophical confusions. Its reliance on extra dimensions, which are not practically observable, raises questions about the physical reality of these dimensions.\footnote{See \cite{Dawid2007, Woit} for more profound philosophical points of view toward the consequences of string theory.} Furthermore, the anthropic principle often invoked to explain why our universe has the particular physical constants it does within the string theory landscape has been criticized for being unscientific, as it resorts to observational selection effects rather than predictive power.\footnote{See \cite{Susskind2003, Weinstein, Cirkovic} and the references therein.} 

\par Considering the aforementioned defects and shortcomings, it is reasonable to question whether our current interpretation of the fundamental concepts of string theory is correct. The persistent lack of empirical evidence, combined with the crucial problems of mathematical complexity, predictive power, and philosophical interpretation, suggests that our prevailing understanding of string theory may be fundamentally flawed. Additionally, the incorrect predictions about the cosmological constant's sign and the dimensions of spacetime further cast doubt on the theory's interpretation's validity. Perhaps the spacetime fields we interpret as the dimensions where strings are embedded do not accurately represent the true nature of spacetime. It may be that an alternative framework of concepts or a deeper understanding of string theory itself is required to resolve these issues and provide a more coherent and empirically verifiable description of the universe.

\par Does string theory really discuss the quantum mechanical fluctuations of strings in spacetime? But if this is not the case, then what would be the inherent meaning of string theory mathematics and solutions? In fact, we have shown rigorously that if a specific collection of quantum states can be attributed to a subunit of a neuron, then the Brownian motions of these quantum states and their interactions in the corresponding Hilbert space will be thoroughly described by the Polyakov path-integral formula. However, we must insist that this beautiful mathematical theory will be well-defined when the number of these quantum states of the neuronal subunit is exactly equal to 25 in order to guarantee the consistency of bosonic string theory in 26 dimensions. Hence, the above calculations can be a decisive defense of string theory against false or unreliable interpretations if we can find neural subunits with 25 independent real quantum states to which the solutions of string theory are applicable. 

\par Therefore, the incompatibility of string theory with laboratory observations and empirical aspects is not, in fact, a flaw in this spectacular, mathematically elaborated quantum theory, but rather our misunderstanding of its fundamental ingredients and underlying theoretical mechanism. In other words, the list of problems attributed to string theory is wrongly placed, as their roots actually stem from our misunderstanding of the objectives, formalisms, theoretical aspects, and contents of string theory. Furthermore, our studies demonstrate that not only do the fundamental formulations of bosonic string theory correctly describe the Brownian motion of quantum states on strings, but they also miraculously include the non-local quantum mechanical interactions inherent in the genuine formula of the Wiener path-integral measure with the highest possible accuracy.


\par
\section{Speculation: String Theory as a Quantum Model of the Brain}
\setcounter{equation}{0} 

\par Now, based on what we have theoretically extracted above, we encounter a profoundly challenging question: If string theory is a theory of the Brownian motion of quantum states on any arbitrary one-dimensional object, can it be a quantum theory for quantum neurobiological effects due to neurons or some of their specific stringy subunits? Or equivalently: Can the results of string theory, which are inherently distinct from laboratory observations, be related to the quantum mind and the quantum mechanical aspects of neurobiological phenomena? Therefore, it is likely that the achievements of string theory regarding the description of the fundamental forces of nature are actually correlated to our mental analysis of the environment, including social/psychological interactions, linguistic relations, or physical events.

\par Based on this viewpoint, string theory could be regarded as the fundamental computational setting for the possible quantum computer of the mind.\footnote{See \cite{Kers, Penrose1999} for more related topics.} Consequently, the Standard Model can be considered a bridge between our mental analysis of the existing world (based on human perceptions) as thoroughly described by string theory and what actually happens in nature that is experimentally observed in laboratories. Therefore, if it is possible to attribute a set of 25 quantum states to each special subunit of a brain neuron, then all the solutions and results of string theory should be attributed to a quantum mechanical space (but not the spacetime manifold with or without extra dimensions) wherein quantum neurological, cognitive, psychological, or consciousness interactions of the brain (such as decisions, speculations, or contemplations) occur.

\par In principle, the most significant emergent phenomenon here is consciousness, which plays a crucial role in the fundamental assumptions of quantum mechanics and interacts with the quantum mechanical ingredients in the measurement processes. It seems that consciousness is a complex and multifaceted phenomenon that is not localized to a single part of the brain; instead, it emerges from the integrated activity of multiple brain regions. However, several key areas are particularly important for different aspects of consciousness: \textbf{1)} The \emph{cerebral cortex} or the \emph{prefrontal cortex},\footnote{See for example \cite{Raccah, Zeman, Bart, Kapoor, Pana} and the references therein.} \textbf{2)} The \emph{thalamus},\footnote{See \cite{Deshmukh, Zeman, Min, Aru, Red, Ward} as a comprehensive presentation of the issue.} \textbf{3)} The \emph{brainstem} or the \emph{reticular activating system},\footnote{See \cite{Parvizi, Edlow, Edlow2, Devor, Zeman} for more discussions about the correlation between consciousness and the brainstem.} \textbf{4)} The \emph{parietal cortex},\footnote{See \cite{Afrasiabi, Liu, Taylor2001, Bor, Zare} for more discussions.} \textbf{5)} The \emph{temporal lobe},\footnote{See \cite{Englot, Dheer, Campora, Yousef} for more information.} and \textbf{6)} The \emph{insular cortex}.\footnote{See \cite{Gu, Tiss, Michel} for more details.}

\par It has been speculated that consciousness arises from the dynamic interactions among these regions, supported by complex networks of neural connections. Modern theories of consciousness, such as the \emph{Global Workspace Theory} (GWT),\footnote{See \cite{Baars1, Baars2} as the pioneering articles of the theory. See also \cite{Finkel, Maille} for more discussions.} and \emph{Integrated Information Theory} (IIT),\footnote{See \cite{Tononi, Haun, Tononi0}.} suggest that it is the widespread and coordinated activity across these networks that gives rise to conscious experience. In his notable book \emph{"The Emperor's New Mind"} \cite{Penrose1999} and later invaluable works such as \emph{"Shadows of the Mind"} \cite{Penrose1995}, Sir Roger Penrose, along with his colleague Stuart Hameroff, an anesthesiologist, speculated that consciousness arises from some quantum processes relating to the basic structures of spacetime and quantum gravity within the brain \cite{Penrose1996}. Specifically, they proposed that \emph{microtubules} within neurons are the key sites where these quantum processes occur. Microtubules are components of the cytoskeleton, providing structural support to cells and playing roles in intracellular transport, cell division, and other essential cellular functions.\footnote{See \cite{Dent} for more information about the neurobiological role of microtubules.}  

\par
\begin{figure}[htbp]
\centerline{\includegraphics{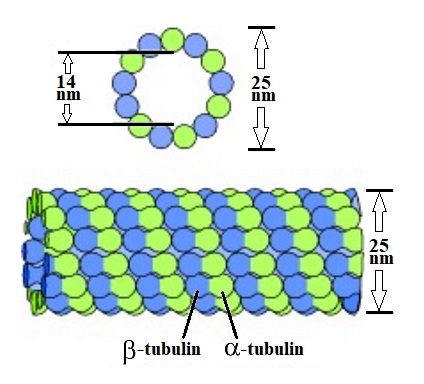}}
\caption{Structure of microtubules: Dimers of $\alpha$- and $\beta$-tubulin polymerize to form microtubules, which are composed of 13 protofilaments assembled around a hollow core \cite{Cooper}.}
\label{fig}
\end{figure}

 According to Geoffrey Cooper \cite{Cooper}, \emph{"microtubules are composed of a single type of globular protein, called tubulin. Tubulin is a dimer consisting of two closely related 55-kDa polypeptides, $\alpha$-tubulin and $\beta$-tubulin. Like actin, both $\alpha$- and $\beta$-tubulin are encoded by small families of related genes. In addition, a third type of tubulin ($\gamma$-tubulin) is specifically localized to the centrosome, where it plays a critical role in initiating microtubule assembly. Tubulin dimers polymerize to form microtubules, which generally consist of 13 linear protofilaments assembled around a hollow core (FIG. \ref{fig}). The protofilaments, which are composed of head-to-tail arrays of tubulin dimers, are arranged in parallel. Consequently, microtubules (like actin filaments) are polar structures with two distinct ends: a fast-growing plus end and a slow-growing minus end. This polarity is an important consideration in determining the direction of movement along microtubules, just as the polarity of actin filaments defines the direction of myosin movement."}\footnote{Although most of the microtubules contain 13 protofilaments, microtubules with various numbers of protofilaments ranging from 8 to 20 have been observed in vitro and in vivo \cite{Sui}.}

\par A pseudo-helical structure is created by the protofilaments joining together laterally. One turn of the helix has 13 tubulin dimers, each originating from a distinct protofilament. Due to the helicity of the turn, in the most prevalent layout, the so-called \emph{"13-3" architecture}, the 13th tubulin dimer interacts with the next tubulin dimer with a vertical offset of 3 tubulin monomers \cite{Sui}.\footnote{Other alternate architectures that have been found at even lower occurrence frequencies are "11-3", "12-3", "14-3", "15-4", and "16-4" \cite{Sui}. In addition, other rare forms of microtubules such as helical filaments have been observed in protist organisms like foraminifera \cite{Bassen}.} There are two possible types of interactions between the subunits of lateral protofilaments within the microtubule, the so-called \textbf{A}-type and \textbf{B}-type lattices. In the \textbf{A}-type lattice, the lateral interactions happen between adjacent $\alpha$- and $\beta$-tubulins (i.e. an $\alpha$-tubulin interacts with a $\beta$-tubulin of the neighboring protofilament). In the \textbf{B}-type lattice, the $\alpha$- and $\beta$-tubulins of protofilament respectively join with the $\alpha$- and $\beta$-tubulins of the neighboring protofilament. Experimental observations have demonstrated that the primary arrangement within microtubules is of \textbf{B}-type. Nevertheless, in most microtubules, there is a seam in which $\alpha / \beta$ tubulins interact \cite{Nogales}.

\par Let us take an overall examination of the molecular structures of microtubules and endeavor to formulate a schematic quantum mechanical model for them. Any standard microtubule comprises 13 protofilaments each of which a complex Schrodinger wavefunction is attributed to. Therefore, each microtubule has $D=26$ real functions, representing the foundational elements of the system's quantum mechanical dynamics. We show them as $X^i$ for $i=1,\cdots, 26$. In other words, the Schrodinger wavefunction $\psi^k = X^{2k-1}+i X^{2k}$ is attributed to the $k$-th protofilament for any $1\leq k \leq 13$. Each Schrodinger wavefunction $\psi^k$ is assumed to be the restriction of an overall Schrodinger wavefunction on the microtubule, say $\widetilde{\Psi}$, to the corresponding protofilament. However, it can be seen that if we consider the topological structure of a microtubule as a cylinder with a specific architecture (e.g. "13-3") and a certain lattice type (e.g. \textbf{B}-type) then all $\psi^k$s cannot be independent.

\par We can simulate a microtubule as a two-dimensional lattice in the ribbon of $[0, N] \times \mathbb{R}$, where $N$ is the number of included tubulins in the protofilaments, and if we consider the \textbf{B}-type lattice structure we can simply assume that each point with integer coordinates represents a specific tubulin.\footnote{For a \textbf{A}-type lattice structure one must shift the vertices of the next horizontal axis of the lattice a fraction of the unity. The amount of this shift depends on more accurate information about the molecular structure of the microtubule. It appears that it causes some complexity in the lattice-based model.} Due to the cylindrical topology of a microtubule, we must identify $(x, y)$ with $(x,y+13)$ in the ribbon. Hence, the $k$-th protofilament is identified by horizontal axes with $y=k$ mod 13. On the other hand, as mentioned above, the molecular features of microtubulin result in $\widetilde{\Psi}$ being entirely determined by the 13 wavefunctions $\psi^k$ ($k = 1, \cdots, 13$) of the 13 protofilaments, which serve as the system's building blocks. Hence, it appears that having knowledge of $X^i$, for $1\leq i \leq 26$, is sufficient to understand the quantum activities of the microtubule.

\par However, one needs to know the number of independent fields $X^i$ to determine the essential quantity of fundamental fields required for analyzing the quantum behaviors of microtubules. We should note that the architectural structure of the microtubule imposes an additional condition on $\widetilde \Psi(x,y)$. More precisely, in the "13-3" architecture, we must assume specific relationships between $\widetilde \Psi(x, y + 13)$ and $\widetilde \Psi(x + 3, y)$. Based on these topological and architectural features, we must introduce a constraint on the fundamental ingredients of the model, the wavefunctions $\psi^k$, $k=1, \cdots, 13$. This reduces the necessary independent functions $X^i$ to 25 out of 26 to determine the Schrodinger wavefunction $\widetilde \Psi$. As we have demonstrated above, the Brownian motion of these independent quantum states will produce the celebrated Polyakov path-integral formulation of the well-defined bosonic string theory in $D = 26$ dimensions, with one augmented non-locality field $X^0$ which encapsulates the non-local effects appearing in the Wiener fractal measure.

\par Therefore, regardless of its range of validity for describing the fundamental forces of nature, the bosonic string theory is the most accurate theoretical model for the quantum mechanical functions of microtubules. Consequently, upon the significance of microtubules in the biological/cellular processes of neurons, the quantum mechanical model of neurons' actions is expressed by the two-dimensional conformal field theory of bosonic strings at the utmost approximation. The underlying assumptions for the mentioned conclusion, including identifying the 13 protofilaments within the main body of a microtubule and modeling a microtubule as a string are theoretically justifiable approximations, since the length of a microtubule may elongate up to 40 microns or more  \cite{Spreng, Yu}, while its diameter is about 25 nanometers \cite{Cooper}. Hence, the ratio of "length/diameter" for a medium microtubule is more than 1600 which is an admissible ratio for the stringy approximation of a microtubule and the identification of protofilaments hypothesis.

\par According to Penrose and Hameroff's \emph{Orch-OR theory} (Orchestrated Objective Reduction theory) microtubules within neurons can support quantum coherence, a state in which quantum particles (like electrons or photons) exist in a superposition of states.\footnote{See \cite{Hameroff2003, Hameroff2012, Hameroff2014, Hameroff2014a, Penrose2014, Hameroff2020, Hameroff2022, Ham-Pen1996a} for more detailed studies on Penrose and Hammeroff's Orch-OR theory.} In principle, these coherent quantum states are suggested to be essential for consciousness. The theory posits that quantum computations occur within the microtubules, and these computations are orchestrated by biological processes \cite{Hameroff2012}. In fact, upon Orch-OR theory consciousness is proposed to emerge when a quantum state within the microtubules undergoes objective reduction (collapse), influenced by the brain's environment and processes. Furthermore, the theory suggests that these quantum events are integrated with classical neural processes, potentially explaining how large-scale brain activity and conscious experience are linked.\footnote{This hypothesis remains highly controversial and speculative within the scientific community. In fact, most neuroscientists and physicists are skeptical of the idea that quantum processes play a direct role in consciousness, largely because of the challenges in maintaining quantum coherence in the warm, wet, and noisy environment of the brain \cite{Tegmark, McKemmish, Reimers1, Reimers2, Villatoro, Baars2012, Georgiev, Litt}. Nonetheless, Penrose and Hameroff's theory has sparked considerable debate and interest in exploring new but mesmerizing interdisciplinary approaches to understanding consciousness.}

\par What we have shown in this study is not, in fact, a theoretical proof of Penrose and Hamroff's theory details, but it demonstrates it is likely that Hameroff and Penrose have correctly predicted the source of the quantum activities of the brain's neurons.\footnote{See \cite{Jedlicka, Roth, Dey, Kher, Donald, Geor, Nan, Koch, Schwartz, Swan} as the other investigations and theories about the quantum mechanical source of consciousness and quantum activities of neurons in quantum neurobiology. In fact, what Nanopoulos has investigated in his seminal article \cite{Nan} is extremely close to the theoretical results of our studies in the present work.} In addition, the special structure of microtubules which comprise 13 protofilaments ensures that the resulting bosonic string theory is consistent and hence, according to Penrose and Hameroff can support quantum coherence. Indeed, there are many known cellular subunits with string-like structures in biology (e.g., substructures of chromatin or the substantial structure of DNA), but it seems that the only one that fulfills the mandatory conditions of a consistent bosonic string theory (having $D-1=25$ real quantum states on a unified stringy structure) is microtubule.

\par On the other hand, the essential role of $\gamma$-tubulins in the assembling of microtubules, and the significance of the inherent polar structure of a microtubule in growing (sewing) of the corresponding stringy structure, as explained by Cooper \cite{Cooper} is surprisingly the reminiscent of strings interactions. In fact, such processes exactly align with the topological interactions of bosonic strings within the Polyakov path-integral mechanism. Therefore, it seems that even the cell division that takes place with the direct and notable involvement of microtubules is actually a biological phenomenon supported by some quantum processes based on the interactions of the bosonic strings of microtubules. However, we have to mention that at present, our understanding of the importance and the role of string theory solutions in explaining biological processes that occur with the involvement of microtubules is not complete and it needs more extensive investigations.

\par Furthermore, since the field $X^0$ encapsulates the non-local effects of the fields $X^i$, $1\leq i \leq 25$, it could theoretically be the source of non-local neurobiological effects within the brain, the large-scale brain activities, and hence the conscious experience. This interpretation aligns with the idea that neuronal quantum interactions may not be limited to classical local exchanges but could also involve quantum mechanical non-locality \cite{Penrose1996}. Consequently, it would be reasonable to refer to the non-locality field $X^0$ as the \emph{consciousness field} in the context of brain activities. Although it may seem that identifying the consciousness field $X^0$ with the time dimension in the string theory formalism is a fundamental mistake, the derivation of the string spectrum in various ways including the use of the light-cone gauge guarantees the practical and theoretical validity of this interpretation. This issue raises a deep understanding of the concept of time and its connection with the fundamental ingredients of consciousness.

\par Meanwhile, in his recent work \emph{"Consciousness Is Quantum State Reduction Which Creates the Flow of Time"} \cite{Hameroff2024}, Hameroff has beautifully speculated based on the contents of Orch-Or theory that the consciousness is the main responsible for the flow of time. Such an exciting result once again shows how close the achievements of our study based on the Brownian formulation of bosonic string theory are to the results of Penrose and Hameroff's theory. In this regard, the role of spacetime curvature in Penrose's view of consciousness can be related or transferred to the interaction of string theory's gravitons as spin 2 bosonic string states with massive string states at low energy levels. Nevertheless, these ingredients inherently involve the excited quantum states of microtubules and bring back the issue of the role of spacetime curvature in consciousness to the same brain quantum activity. Strictly speaking, it seems that Penrose's objective reduction is also the inevitable consequence of the string-theoretic interactions of microtubules. This point not only recalls the role of microtubules in orchestrating the objective reduction process in the Orch-OR theory but may refer the objective reduction process to the microtubules themselves.

\par As the most pivotal ingredient of the Orch-OR theory and our achievements of string-theoretic formulation of the brain quantum activities, the microtubules are \emph{time crystals} \cite{Ghosh}, and according to Hameroff, it appears a microtubule-based scale-invariant hierarchy exists in the brain and grows inwardly to include quantum-level phenomena with non-locality \cite{Hameroff2020, Hameroff2024}. Therefore, it can be seen that producing a consistent string theory by quantum fluctuations of microtubules is in remarkable agreement with Penrose and Hameroff's theory. Although this alignment still does not mean to prove the Orch-OR theory or its coincidence with our Wiener-Brownian formulation of bosonic string theory, the calculations presented above provide a spectacular mathematical framework for studying this theory. The discussion about the connection between Penrose and Hameroff's theory and the achievements of string theory within the unprecedented interpretations we present here requires more extensive separate investigations and is deferred to future studies.

\par Regardless of the possible correlation between the Orch-Or theory and our Brownian interpretation of bosonic string theory for microtubules, both theories fundamentally include non-local quantum mechanical interactions and address it as the fundamental source of consciousness. The consciousness field $X^0$ may contain the whole non-local features of consciousness in Penrose-Hameroff's theory, but its main role in the string-theoretic formulation of microtubules is to drag the whole non-local quantum effects of the Wiener fractal measure into the final formulation. While this provides a fascinating insight into the non-local processes in the brain, it also opens the door to cautious speculation: If telepathy has any scientific basis, the consciousness field $X^0$ might be a candidate for facilitating such a phenomenon via certain quantum mechanical mechanisms.\footnote{However, this remains speculative and requires rigorous empirical validation before drawing any definitive conclusions. See also \cite{Brassard1999, Brassard2003, Brassard2005, Gisin2007, Xu2022} for more similar, and perhaps correlated discussions about the relation of quantum pseudo-telepathy and quantum entanglement.}


\par
\section{Summary and Conclusion}
\setcounter{equation}{0}

\par In our study, our primary aim was to investigate the Brownian motion of several real quantum states along a string utilizing the Wiener fractal measure we introduced in \cite{Varshovi1}. Our findings revealed that by excluding the non-local contributions in the Wiener measure, the resultant formulation aligns with the Polyakov path-integral measure in Euclidean spacetime (Section II). In our investigation, we found that reintroducing non-local effects through the incorporation of an augmented real state $X^0$, the so-called non-locality field, led to a significant observation: the Wiener fractal measure yields the Polyakov path-integral measure of bosonic string theory with the utmost accuracy (Section III).

\par Although this result was surprising in its own turn, the distinct essence of the non-locality field $X^0$ with its inherent substance in string theory raised a profound question in our understanding and interpretation of the fundamental ingredients, significant principles, and substantial concepts of string theory, reshaping our understanding and interpretation of this complex field. This prompted a thorough reassessment of string theory's experimental implications and the discrepancies between its theoretical predictions and empirical data. It also raised the intriguing possibility that string theory may be addressing concepts beyond the quantum dynamics of strings, which are hypothesized to underlie the fundamental forces of nature (Section IV).

\par Furthermore, the conformation of the ultimate expression for the Brownian motion of quantum states on strings and the Polyakov path-integral formulation, reinforces this perspective, since our analysis delved into the stochastic behavior of quantum fields associated with any string, regardless of its fundamental role in natural forces. Hence, our endeavor focused on identifying a cellular component within brain neurons possessing a molecular configuration resembling a string so that this sought-after entity serves as a tangible illustration of the natural manifestation of string theory principles (Section V). This quantum machine likely plays a significant role in mediating essential quantum interactions within the brain.

\par Fortunately, Penrose and Hameroff were able to identify these quantum machines within the brain approximately three decades ago: Microtubules \cite{Ham-Pen1996a, Penrose1996}. Microtubules are essential cellular subunits involved in critical biological processes, such as cell division. Also, they comprise 13 protofilaments, each of which represents a natural example of the fundamental elements of string theory based on our calculation in Section III. Attributing a complex Schrodinger wavefunction to each protofilament results in 26 real functions, encompassing 25 independent fields $X^i$ ($1 \leq i \leq 25$), due to the topology and architecture of the microtubule. Considering the non-locality field $X^0$, we find that the Brownian motion of quantum states on microtubules manifests as a coherent and sustainable bosonic string theory in $D=26$ dimensions within brain neurons (Section V).

\par Consequently, our achievements in this paper not only confirm the main parts of the Orch-OR theory, but they provide a computational framework for the quantum activities of the brain. In addition, our mathematical formulation of bosonic string theory in the framework of the Wiener fractal measure and in the presence of the non-locality field offers a coherent structure for comprehending Penrose and Hameroff's theory fundamental concept: Consciousness. We further argued that the role of consciousness must be attributed to the non-locality field $X^0$, which is referred to here as the consciousness field. This conclusion is in complete agreement with Hameroff's recent finding that identifies consciousness and time \cite{Hameroff2024}.


\section{Acknowledgments}

\par This article is fervently dedicated to the cherished memory of Mohammad Reza Pahlavi and his illustrious father, Reza Pahlavi, whose legacies continue to inspire and resonate deeply within our hearts. In addition, the author wishes to extend his heartfelt gratitude to Dr. Bahram Ghiassee, a distinguished visiting academic at the University of Surrey. His invaluable assistance and insightful guidance have been a beacon of inspiration and support throughout the author's study.





\end{document}